\documentclass[superscriptaddress,prl,aps,longbibliography,twocolumn]{revtex4-2}
\usepackage{tikz}  
\usetikzlibrary{shapes,arrows,positioning,automata,backgrounds,calc,er,patterns}
\usepackage{graphicx}
\usepackage{dcolumn}
\usepackage{bm}
\usepackage{amsmath}
\usepackage{amssymb}
\usepackage{amsthm}
\usepackage{amsfonts}
\usepackage{enumerate}
\usepackage{array}
\usepackage{changes}
\newcolumntype{P}[1]{>{\centering\arraybackslash}p{#1}}
\usepackage{float}
\usepackage{dsfont}

\newcommand{\ket}[1]{\left|#1\right\rangle}
\newcommand{\bra}[1]{\left\langle#1\right|}

\newcommand{\rev}[1]{\textcolor{black}{#1}}

\usepackage[colorlinks=true,urlcolor=blue,citecolor=blue,allcolors=teal]{hyperref}
\usepackage{cleveref}

\begin{document}

\title{Optimal quantum \rev{reservoir} learning in proximity to universality}

\newcommand{\Aalto}{MSP Group, Department of Applied Physics, Aalto University, FI-00076 Aalto, Espoo, Finland}
\newcommand{\ICMM}{Interdisciplinary Centre for Mathematical Modelling and Department of Mathematical Sciences,\\ Loughborough University, Loughborough, Leicestershire LE11 3TU, United Kingdom}
\newcommand{\LUPhys}{Department of Physics, Loughborough University, Loughborough, LE11 3TU, United Kingdom}

\author{Moein N. Ivaki}
\email[\vspace{-3pt}]{moein.najafiivaki@aalto.fi}
\address{\Aalto}
\author{Matias Karjula}
\address{\Aalto}
\author{Tapio Ala-Nissila}
\email[\vspace{-3pt}]{tapio.ala-nissila@aalto.fi}
\address{\Aalto}
\address{\ICMM}

\date{\today}

\begin{abstract}

\rev{The study of the boundary between classically simulable and computationally complex quantum dynamics is fundamental to understanding which physical resources may enable enhanced information-processing capabilities.} We investigate this within the framework of quantum reservoir computing by introducing a tunable $N$-qubit random circuit model, where a fraction $p$ of Clifford gates are probabilistically substituted with nonstabilizing conditional-$\hat{T}$ gates. We establish a direct correspondence between the reservoir’s performance on temporal processing tasks and its entanglement spectrum statistics and long-range nonstabilizer resource content. 
To assess scalability, we study the scaling of the anti-flatness of states in the large-$N$ limit at a fixed circuit depth ratio \(d/N \sim \mathcal{O}(1)\). This is taken as a witness to concentration of measures, a known impediment to learning in thermalizing systems. We demonstrate that the learnability and scalability of the reservoir can be continuously controlled by the parameter $p$, allowing us to navigate from classically tractable to maximally expressive quantum dynamics. \rev{These architecture-agnostic results provide a general strategy for designing tunable and expressive quantum reservoirs, highlighting how certain nonclassical properties control average-case intrinsic learnability and functionality.}

\end{abstract}
\maketitle

\textbf{\emph{Introduction.---}}Quantum machine learning (QML) is among the most significant near-term applications of quantum devices and computing. The cornerstone in QML circuit design is  the variational parametrization of quantum circuits~\cite{cerezo2022challenges}, analogous to classical neural networks. However, such approaches suffer from serious trainability problems \cite{mcclean2018barren,larocca2025barren,de2025behind,deshpande2024dynamic}. Despite efforts to remedy the scalability issues, no satisfactory solution has been found \cite{cerezo2023does,cerezo2021cost, bermejo2024quantum, lh6x-7rc3}. This has led to considering ``post-variational" QML \rev{(PVQML)} \cite{huang2023post}, where {fixed} but sufficiently {expressive} random dynamics are exploited \cite{nakajima2021reservoir, mujal2021opportunities}. A central challenge with regard to this is to design protocols that exceed the classically simulable limit in a controllable way while avoiding known pathologies~\cite{thanasilp2024exponential, xiong2025fundamental}. 

\begin{figure}[t]
    \centering
    \includegraphics[width=0.77\linewidth]{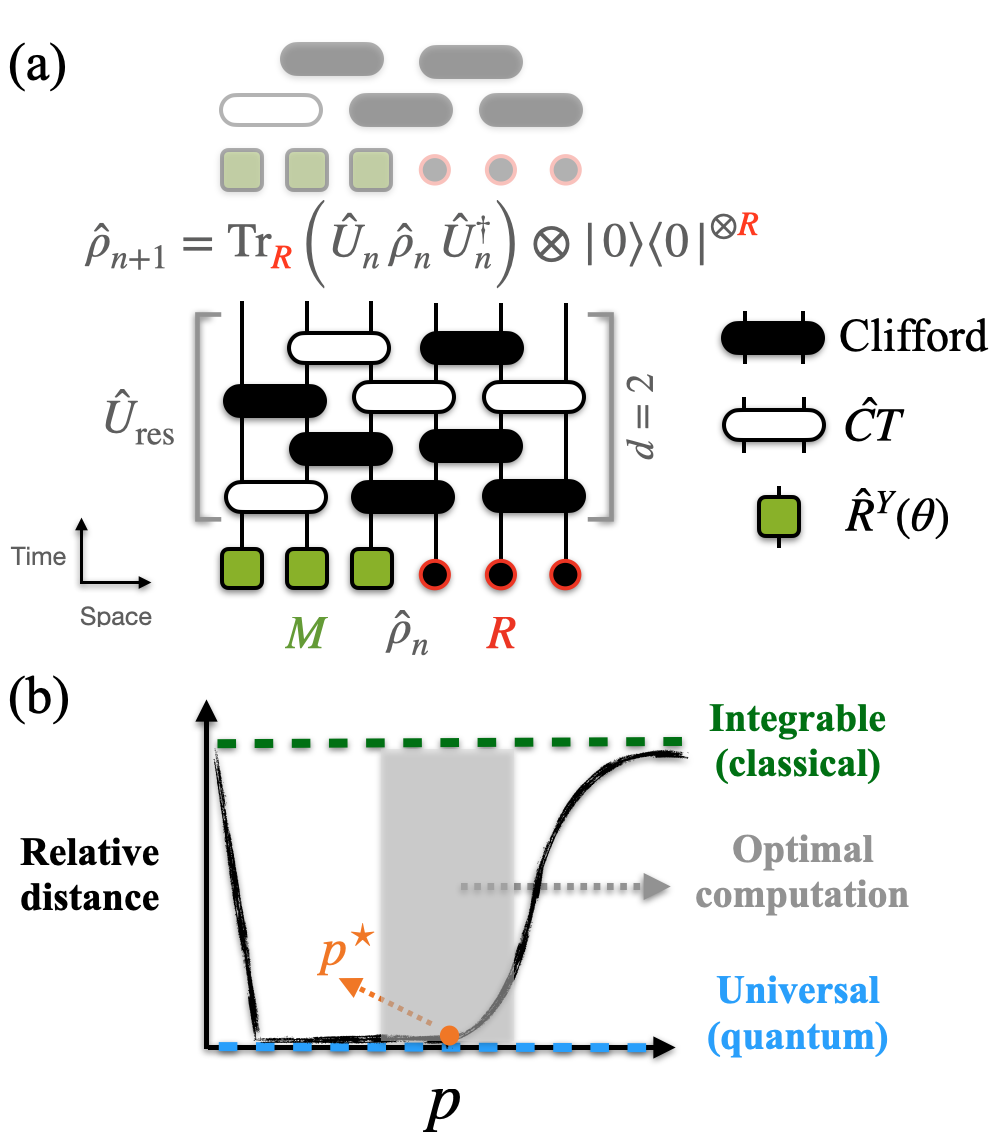}
    \caption{\textbf{A probabilistic quantum reservoir computer.} \textbf{(a)} The circuit’s qubits are split into memory $M$ and readout $R$ subsets, and evolution is captured by an effective contractive channel with the unitary update $ \hat{U}_n\equiv\hat{U}_{\rm res}\,\bigotimes_{j\in M} \hat{R}^{Y}_j(\theta_n)$, followed by (idealized) measure-and-reset of readout qubits. Classical inputs, $\theta_n$, are encoded as local rotations $\hat{R}^Y(\theta):=\exp [-i\hat{Y}\theta/2]$, and reservoir is iterated for $n$ steps. \textbf{(b)} \rev{Average} learnability is characterized via \rev{\textit{relative distance to a quantum universal entanglement spectrum}, with $p^{\star}(d)$ indicating the depth-dependent onset of a chaotic-integrable crossover.} In an optimal $(p,d)$ window, a linear readout on the model’s observables accurately reconstructs nonlinear, time-dependent functionals of the input history. \rev{Throughout, ``optimal" denotes average-case intrinsic learnability, not task-driven model optimization.} }
    \label{fig1:scheme}
\end{figure}

From the Gottesman-Knill theorem it follows that universal quantum computing requires finite ``magic", {\it i.e.}, resources beyond the Clifford gates, stabilizer states and Pauli measurements~\cite{nielsen2010quantum,harrow2017quantum,PhysRevA.71.022316}. While it is well-known that non-Clifford resources are an essential ingredient to unitary designs (pseudo-randomness) and genuine quantum chaotic behavior~\cite{leone2021quantum, webb2015clifford, mele2024introduction}, \rev{much less is known regarding how such nonclassical resources shape the performance of PVQML algorithms.} Despite attempts to correlate learnability to entanglement, coherence, and scrambling~\cite{PhysRevLett.124.200504,PhysRevResearch.3.L032057,PhysRevLett.126.190501, palacios2024role, PhysRevResearch.3.033090,PRXQuantum.2.040316}, these metrics alone cannot reliably predict performance or viability. \rev{In PVQML, particularly quantum reservoir computing,} one commonly employs many-body Hamiltonians or random circuits where the influence of distinct quantum resources is inseparable~\cite{PhysRevApplied.8.024030, PRXQuantum.5.040325, hu2024overcoming, mccaul2025minimal, wzwv-7rk2, vetrano2025state}, obscuring causal role in temporal learning. Rather than directly addressing quantum advantage~\cite{huang2025vast,PRXQuantum.3.030101}, here we aim to analyze how nonlocal physical resources structure learnability and functionality of quantum reservoirs \rev{(QRs)}.

In this work, we propose a paradigm of {\it probabilistic QRs} (PQRs) that interpolate between volume-law-entangled stabilizer and weakly-entangled (localized) extremes, hosting an intermediate dynamical regime exhibiting finite and controllable long-range magic. This is induced via probabilistically replacing random Clifford gates by conditional-$\hat{T}$ gates, $\hat{CT}:=\hat{I}\otimes\ket{0}\bra{0}+\hat{T}\otimes\ket{1}\bra{1}$, $\hat{T}={\rm diag}(1,e^{i\pi/4})$ (Fig.~\ref{fig1:scheme}(a)). We show that similar to $\hat{T}$-doped Clifford circuits, this provides a resource-efficient and methodological route to quantum universality~\cite{li2023optimality,haug2025probing,PRXQuantum.5.030332,PRXQuantum.6.020324, niroula2024phase,PhysRevLett.130.240602}. The replacement paradigm \rev{additionally} allows systematic and controlled generation of entanglement and magic. 

Our main result is that the \rev{ensemble-typical} learning performance of the PQRs is correlated to \emph{entanglement-spectrum complexity}. To unfold this, we consider a measure of entropic divergence for the spectral distribution of subsystem entanglement entropy, and quantify the relative complexity with respect to a known universal distribution. Emphasizing on the role of \emph{mutual magic}, we further find that the ideal (nonlinear) memory and information processing capabilities are reached at the crossover between a quantum chaotic and an integrable regime. Remarkably, in our PQR this is accessed by only tuning the probability $p$ with a fixed circuit depth $d\sim\mathcal{O}(N)$~(Fig.~\ref{fig1:scheme}(b)). We also show that concentration of measures, typical of scrambling quantum dynamics, can be inferred from the finite-size scaling of \textit{anti-flatness} of states. We relate this self-sampling variance to the expected functionality of reservoir computing in large-$N$ limit.

\textbf{\emph{Model.}---}We consider a nearest-neighbor brickwork circuit of fixed depth $d$ and $N$ qubits. The local probabilistic unitary map is  \[\hat\rho\mapsto \begin{cases} \hat{CT}\,\hat\rho\,\hat{CT}^\dagger, & \text{w.p. } p; \\ \hat{U}_{\mathrm{C}}\,\hat\rho\,\hat{U}_{\mathrm{C}}^\dagger, & \text{w.p. } 1-p, \end{cases}\]where $\hat{U}_{\mathrm{C}}$ is a random Clifford gate and $\hat\rho$ is a quantum state. The spacetime volume is $V=(N-1)\,d$, and total number of $\hat{CT}$ gates is extensive for $p>0$, obeying the binomial distribution with the expected value $\mathbb{E}[\#\hat{CT}]=pV$. The realized density $\#\hat{CT}/V$ then converges to the expected density, $p$, with fluctuations $\sqrt{p(1-p)/V}$. Clifford gates are generated by the set $\{\hat{H}, \hat{S}, \hat{CX}\}$, where $\hat{H}$ is the Hadamard gate $(\hat{X}+\hat{Z})/\sqrt{2}$, $\hat{S}=\sqrt{\hat{Z}}$ and $\hat{CX}= \hat{I}\otimes\ket{0}\bra{0}+\hat{X}\otimes\ket{1}\bra{1}$~\cite{9435351}. These gates conjugate the $N$-qubit Pauli group $\mathcal{P}_N=\{\hat{I},\hat{X},\hat{Y},\hat{Z}\}^{{\otimes}N}$ to itself (up to a phase); {\it i.e.}, for all $\hat{P}\in\mathcal{P}_N$ one has $\hat{U}_\mathrm{C}\,\hat{P}\,\hat{U}_\mathrm{C}^{\dagger}\in \mathcal{P}_N$. Since any Pauli string (across any bipartition) stays a tensor product under the action of Clifford group, \textit{operator entanglement} for such string remains \textit{zero} at all times~\cite{fisher2023random, mi2021information}. For a stabilizer state \(\ket{\mathrm{\texttt{stab}}}=\hat{U}_\mathrm{C}\ket{0}^{\otimes N}\)~\cite{gottesman2016surviving}, the reduced density matrix on a subsystem \(R\) (the readout) satisfies \(\hat{\rho}_R \propto P_R\), where $P_R=P_R^2$ is a projector. This implies that the nonzero entanglement-spectrum is highly degenerate and displays no level repulsion. Such properties renders classical simulation of these circuits highly efficient with polynomial resources~\cite{PhysRevA.70.052328}. Interestingly, even a single $\hat{T}$ gate may drive the entanglement-spectrum distribution to a universal Wigner-Dyson form~\cite{10.21468/SciPostPhys.9.6.087} and lift its flatness~\cite{PhysRevA.109.L040401,true2022transitions}. Nevertheless, in the absence of a universal quantum gate set, the distribution alone does not necessarily signify either universality or classical hardness~\cite{projansky2024entanglement}. 

\begin{figure}[t]
    \centering
    \includegraphics[width=0.999\linewidth]{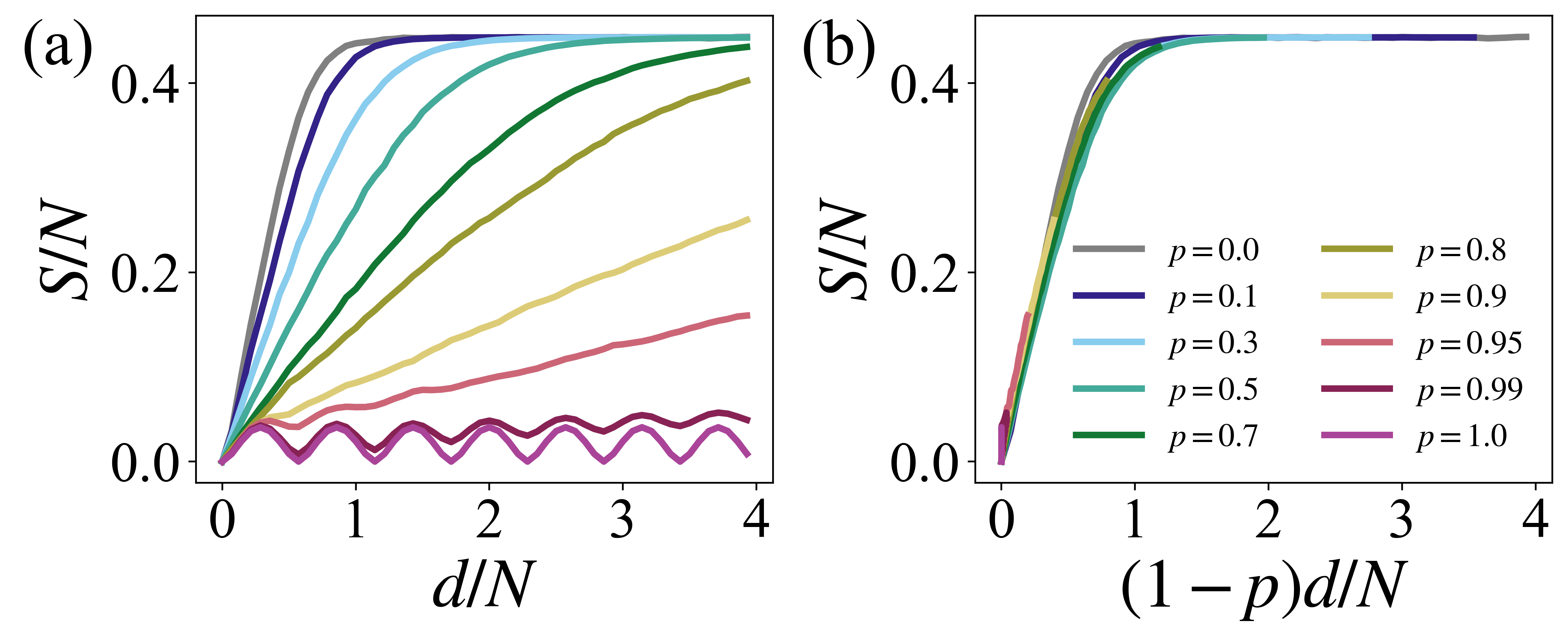}
    \includegraphics[width=0.999\linewidth]{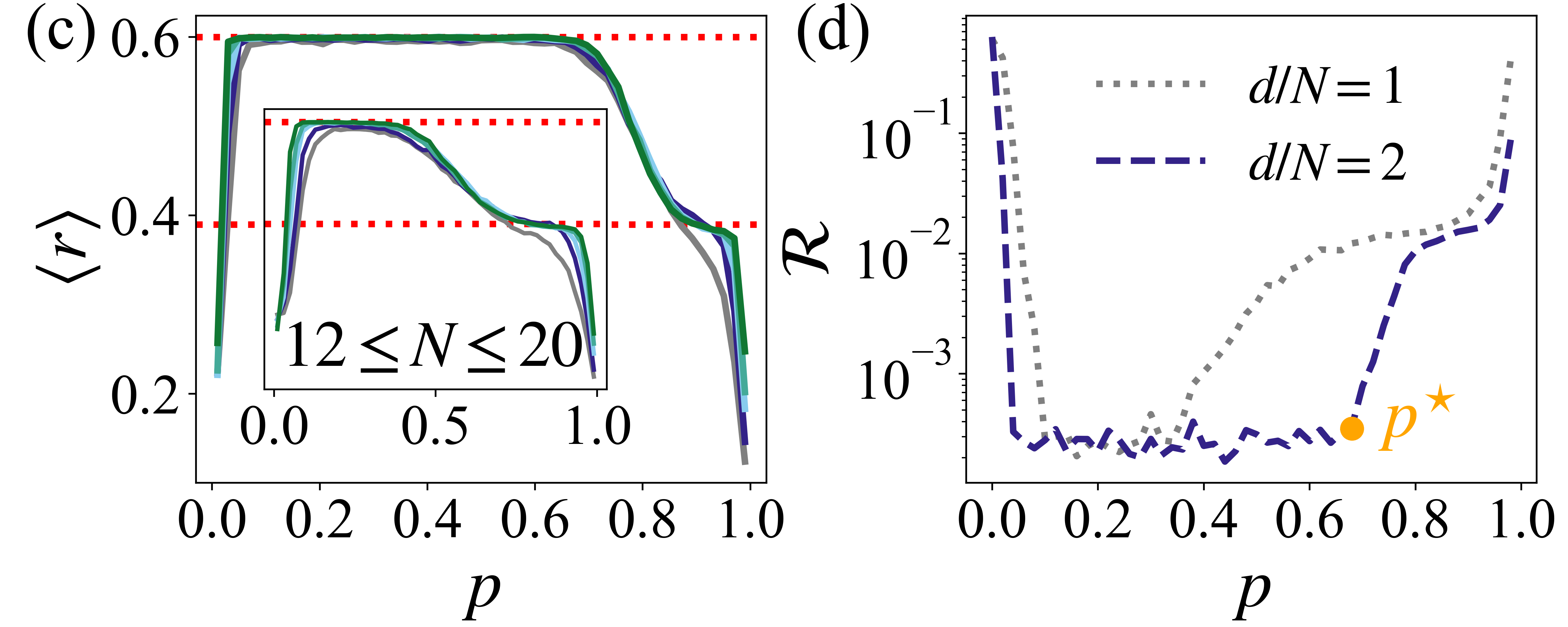}
    \caption{\textbf{Properties of entanglement.} \textbf{(a)} Entanglement dynamics starting from a random product-state for $N=14$. \textbf{(b)} Same vs the rescaled depth $(1-p)d/N$. \textbf{(c)} Mean level spacing ratio $\langle r\rangle$ for $d/N=2$ and (inset) $d/N=1$, computed by taking the mean of the neighboring eigenvalue ratios $r$ over the entire spectrum. The dotted lines mark the standard quantum and classical reference limits, characterized by the GUE for quantum chaotic complex Hermitian matrices, and by the Poisson statistics for integrable systems with uncorrelated spectra, where $\langle r_{\mathrm{GUE}}\rangle \approx 0.6$ and $\langle r_{\mathrm{Poisson}}\rangle \approx 0.39$, respectively. Plotted for various $N$ by starting from the initial state $\ket{0}\!\bra{0}^{\otimes N}$. \textbf{(d)} Relative entropy of distributions for $N=20$. $p^{\star}\approx0.65\pm0.05$ denotes the onset of the crossover to integrable (classical) regime for $d/N=2$. The results are averaged over $200-600$ independent realizations.}
    \label{fig2:ent}
\end{figure}

\textbf{\emph{Entanglement spectrum.---}}We now characterize the dynamics of entanglement and its spectral fingerprints in our model. In the following we fix the spatial size of the readout $R$ and memory $M$ to be equal, so that for a measure of state entanglement entropy, say $S_M:={-\mathrm{tr}[\hat{\rho}_M \log \hat{\rho}_M]}$, we have $S_M^{\infty}=S_R^\infty$, where $S^{\infty}\equiv S(d=\infty)$. For \(p=0\) the circuit is Clifford-only, so an initial stabilizer state remains as such while exhibiting volume-law entanglement with a linear initial growth, saturating at depth \(d^\star\sim\mathcal{O}(N)\) ~\cite{PhysRevX.7.031016, PhysRevLett.111.127205}. Shown in Fig.~\ref{fig2:ent}(a,b), when a fraction \(p\) of entangling Clifford gates is replaced by diagonal \(\hat{CT}\) gates, the entanglement front slows down with an approximate effective velocity \(v_{\mathrm E}(p)\propto (1-p)\). This behavior follows from the hierarchy in average entangling power,
$e_{\hat {CT}}\!\ll\! e_{\hat {CX}}$~\cite{PhysRevA.70.052313,PhysRevA.62.030301}, so that in the limit $p\!\to\!1$ entanglement transport is trivially arrested. Evidently, entanglement curves collapse when plotted against the rescaled time \(\sim\! {(1-p)\,d}/{N}\) and saturation depth scales as $d_{\mathrm{sat}}(p)\approx d^{\star}\,(1-p)^{-1}$. Concretely, $S(d,N,p)\approx s^{\infty} \min[v_{E}(p)\,d, N/2]$, where $s^{\infty}$ is the equilibrium entropy density. Since here we take $d\sim\mathcal{O}(N)$, entanglement may remain extensive for any $p\neq1$, implying that the cost of a tensor-network simulation grows as $\sim\mathcal{O}\left(\,\exp\left[\rev{c}(1-p)Ns^{\infty}\right]\right)$~\cite{orus2014practical}, \rev{with $c$ a constant.} 

It is imperative to probe how a subsystem's spectral distribution, $Q$, approaches the prediction of random matrix theory~\cite{PhysRevLett.110.084101}. Let us consider the spectrum of the reduced density matrix via $\hat{\Omega}_{M}:=-\log\mathrm{Tr}_{R}(\hat{\rho})$, and study the ratio $r_i:=\min(\delta_i,\delta_{i+1})/\max(\delta_i,\delta_{i+1})$, where $\delta_{i}:=\epsilon_{i-1}-\epsilon_{i}$, with $\epsilon_{i}\geq \epsilon_{i+1}$ and $\epsilon_{i}\in {\mathrm{Spectra}(\hat{\Omega}_{M})}$~\cite{PhysRevB.75.155111, PhysRevLett.112.240501,Shaffer_2014}. The reference quantum chaotic distribution here is characterized by the Gaussian unitary ensemble $Q_{\mathrm{GUE}}(r)\propto(r+r^2)^2/(1+r+r^2)^{4}$, and the relative entropy $
 \mathcal{R}(Q\Vert Q_{\mathrm{GUE}})
:= \operatorname{Tr}\!\left[Q\log \big(Q/Q_{\mathrm{GUE}}\big)\right]$ is taken as a measure of informational divergence~\cite{kullback1951information}. 

Plotted in Fig.~\ref{fig2:ent}(c), for nearly any $p>0$ the average spacing ratio converges to the value $\langle r\rangle\approx0.6$ set by the reference distribution, highlighting the extreme sensitivity of the Clifford group to nonstabilizing perturbations. Same is reflected in the behavior of the relative entropy $\mathcal{R}$, plotted in Fig.~\ref{fig2:ent}(d). Importantly, at a $d$-dependent probability $p^*(d)$ the system evolves smoothly to a weakly-entangled state, following the decreasing trend of entanglement from its saturation value. At a relatively lower depth $d=N$, features qualitatively stay the same, yet the crossover naturally takes place at a lower $p^*$. Crucially, for large enough $N$, this crossover admits \textit{size-independent statistics}. It is in the vicinity of this chaotic-integrable crossover that an optimal learning regime emerges and the system's memory becomes maximal.

\textbf{\emph{Mutual magic.---}}As a proxy to nonlocal magic, we consider the behavior of the mutual stabilizer R\'enyi entropy (SRE) for complementary subsystems of size $N/2$~\cite{PhysRevLett.128.050402, wang2025magic, tarabunga2025efficient,haug2026efficient,PRXQuantum.4.040317,PhysRevA.111.052443}, defined as 
\begin{align}
    \mathcal{I}(\hat{\rho}):= \mathcal{M}(\hat{\rho})- \mathcal{M}(\hat{\rho}_{M})-\mathcal{M}(\hat{\rho}_{R}).
\end{align}
For a mixed-state $\hat\rho$, the second SRE is expressed as 
\begin{align}
\mathcal{M}(\hat{{\rho}}):=-{\rm log}\left(2^{-N}\sum_{\hat{P}\in\mathcal{P}_N}|{{\rm Tr}(\hat{\rho} \hat{P})}|^4\right)-S_2, 
\end{align}
where $S_2:=-\log(\mathrm{Tr\,}\hat\rho^2)$ is the 2-R\'enyi entropy. In the following we refer to $\mathcal{M(\hat{{\rho}})}$ as total magic (TM) and $\mathcal{I(\hat{{\rho}})}$ as mutual magic (MM). SREs are a measure of spread over the Pauli basis, indicating deviations from stabilizer states, which concentrate all Pauli weight on exactly $2^N$ strings. By additivity, $\mathcal{M}(\hat\rho_M \otimes \hat\rho_R)=\mathcal{M}(\hat\rho_M)+\mathcal{M}(\hat\rho_R)$, and MM vanishes for separable subsystems, while TM can be nonzero or even maximal.
Observing $\mathcal{I}>0$ thus signals correlations across a cut and quantifies the \textit{long-range} magic of \textit{states}, which arises when a unitary map entangles Pauli operators over some spatial distance~\cite{andreadakis2025exact, dowling2025bridging}. Evolution of magic measures can be tracked to identify quantum phase transitions~\cite{falcao2025magic} and study information scrambling~\cite{10.21468/SciPostPhys.16.2.043,varikuti2025impact}. Useful to note that local Haar-random circuits saturate TM at depth $d \propto \ln N$ to the value $\mathcal{M}^{\infty}_{\mathrm{H}}\sim N-2+\mathcal{O}(2^{-N})$~\cite{turkeshi2025magic, y9r6-dx7p, szombathy2024spectral}.

\begin{figure}
    \centering
    \includegraphics[width=0.6\linewidth]{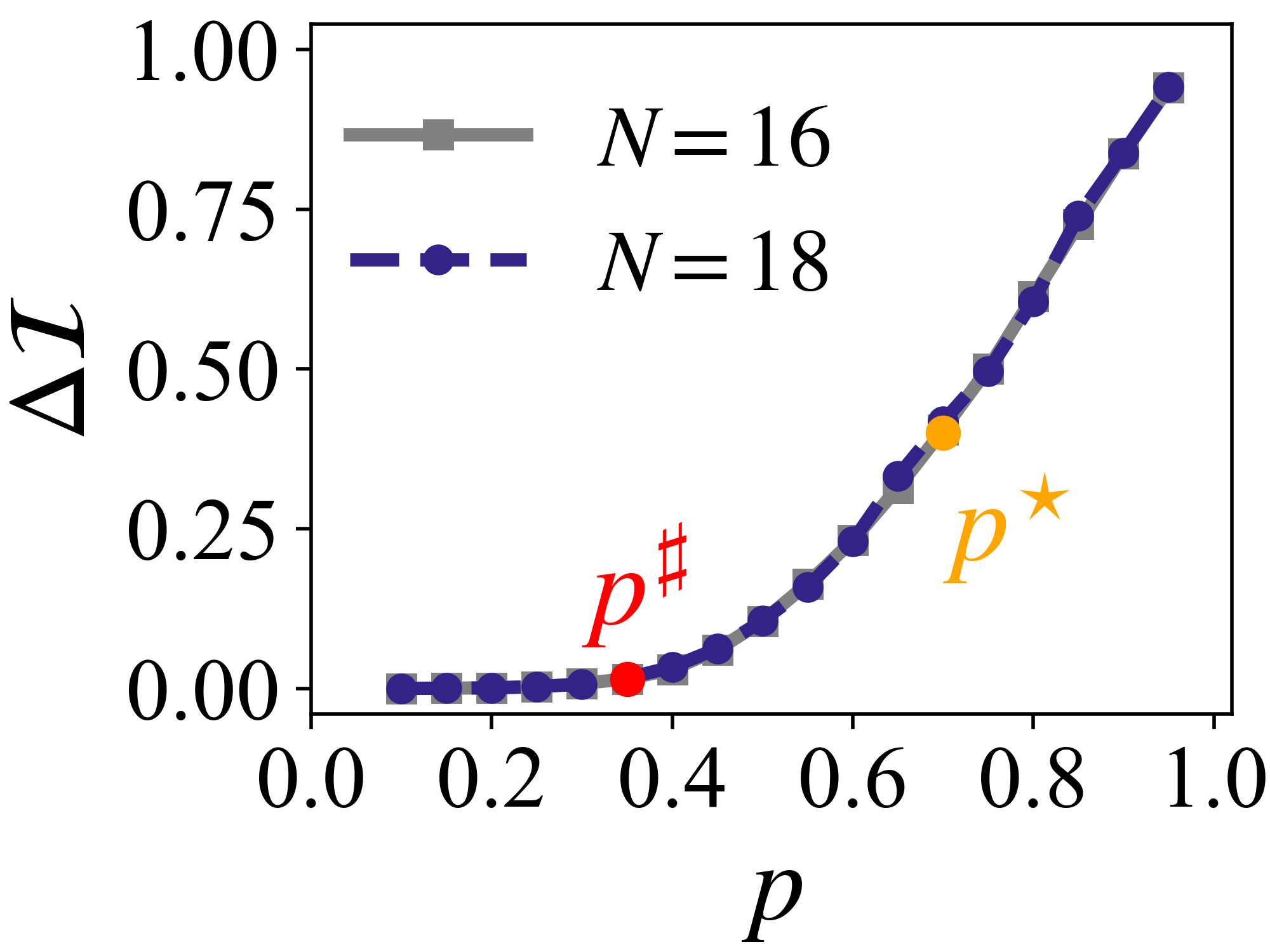}
    \caption{\textbf{Relative gap of mutual magic.} $\Delta\mathcal{I}$ as a function of $p$ at depth $d/N=2$. $p^{\sharp}\approx0.35\pm 0.05$ denotes the point after which, for sufficiency large systems, MM becomes relatively submaximal. The results are averaged over $200$ independent realizations.} 
    \label{fig3:magic}
\end{figure}

To capture the behavior relevant to learnability, we consider the relative gap of MM with respect to Haar random circuits, defined as
\begin{align}
    \Delta\mathcal{I}(p;d)
    = \frac{\big|\mathcal{I}(p;d)-\mathcal{I}_{\mathrm{H}}^{\infty}\big|}{\mathcal{I}^{\infty}_{\mathrm{H}}}.
\end{align}
Intuitively, at a fixed depth $d=2N$ and for $0<p<1$, $\Delta\mathcal{I}(p;d)$ should asymptotically tend to zero as a function of $N$ up to a point $p^{\sharp}$, {\it i.e.} $\Delta\mathcal{I}(p\leq p^{\sharp};2N)\to 0$ as $N\to\infty$ (note that $\mathcal{I}^{\infty}_{\mathrm{H}}\propto N$). For sufficiently large $N$, $\Delta\mathcal{I}$ then should increase with $p$ for $p>p^{\sharp}$. The relative gap is bounded $0\leq\Delta\mathcal{I}\leq1$, and $p^{\sharp}$ marks the onset of a regime where the circuit approaches a non-Haar fixed-point and MM stays submaximal. \rev{This follows since, in the absence of basis-mixing operations, the $\hat{CT}$ gates can become redundant under repetition (up to a finite periodicity), thereby limiting the dynamical growth of non-Clifford complexity.} Concurrently, the resulting state at a fixed depth also becomes progressively less entangled when increasing $p$. 

Fig.~\ref{fig3:magic} shows how the above features are realized in the PQR. One also finds $p^{\sharp}(d)\lessapprox  p^{\star}(d)$, implying a quantum system may exhibit universal entanglement-spectrum statistics without MM or entanglement entropy being necessarily maximal. While closely related, these quantities fundamentally capture different aspects of a quantum chaotic behavior~\cite{leone2021quantum, PhysRevLett.131.180403, PhysRevB.100.125115, PhysRevLett.123.190602, PhysRevX.8.021014, roberts2015localized, PhysRevX.8.021013, jonay2018coarse} (see Appendix). Next we show that learning performance peaks in an intermediate regime where $\mathcal{I}$ is substantial yet submaximal. This suppresses the over-scrambling that degrades learnability, implying that the finite relative gap $\Delta \mathcal{I}\neq0$ can be interpreted as an \rev{accessible} resource for purposeful computations.

\textbf{\emph{Memory and learnability.}---}To perform computations and process the stream of inputs $\{\theta_n\}$, the reservoir is iterated with a fixed random template, and expectation values of diagonal observers $\langle\hat{Z}_i\rangle$, $\langle\hat{Z}_i\hat{Z}_j\rangle_{i\neq j}$ are then recorded at each step to be used for linear regression~\cite{PhysRevApplied.8.024030}. \rev{Learnability metrics} are then averaged over different inputs and random circuits. The effective iterative dynamics is given by $\hat{\rho}_{n+1}={\rm Tr}_{R}\left(\hat{U}_n\hat{\rho}_{n}\hat{U}_n^{\dagger}\right)\otimes|0\rangle\langle 0|^{\otimes {R}}$, where $\hat{U}_n=\hat{U}_{\rm res}\,\bigotimes_{j\in M} \hat{R}^{Y}_j(\theta_n)$ (cf. Fig.~\ref{fig1:scheme}). We assume idealized measurements and reset of readout qubits, with noiseless operations and no restrictions on the number of shots. The general purpose here is to recover a (nonlinear) temporally dependent functionals $y_n=f\left[\{\theta_{n'}\}_{n'\leq n}\right]$ of the input sequence. Employing a simple linear post-processing layer lets us capture the \textit{intrinsic} computational capacity of the PQRs. A suitable reservoir must support a contractive temporal map~\cite{RevModPhys.86.1203, ivaki2025dynamical,PhysRevLett.127.100502, gq9r-d5q8}, which can be diagnosed by the decay of a measure of averaged state-distance (see Appendix). 

\begin{figure}[t!]
    \centering
    \includegraphics[width=\linewidth]{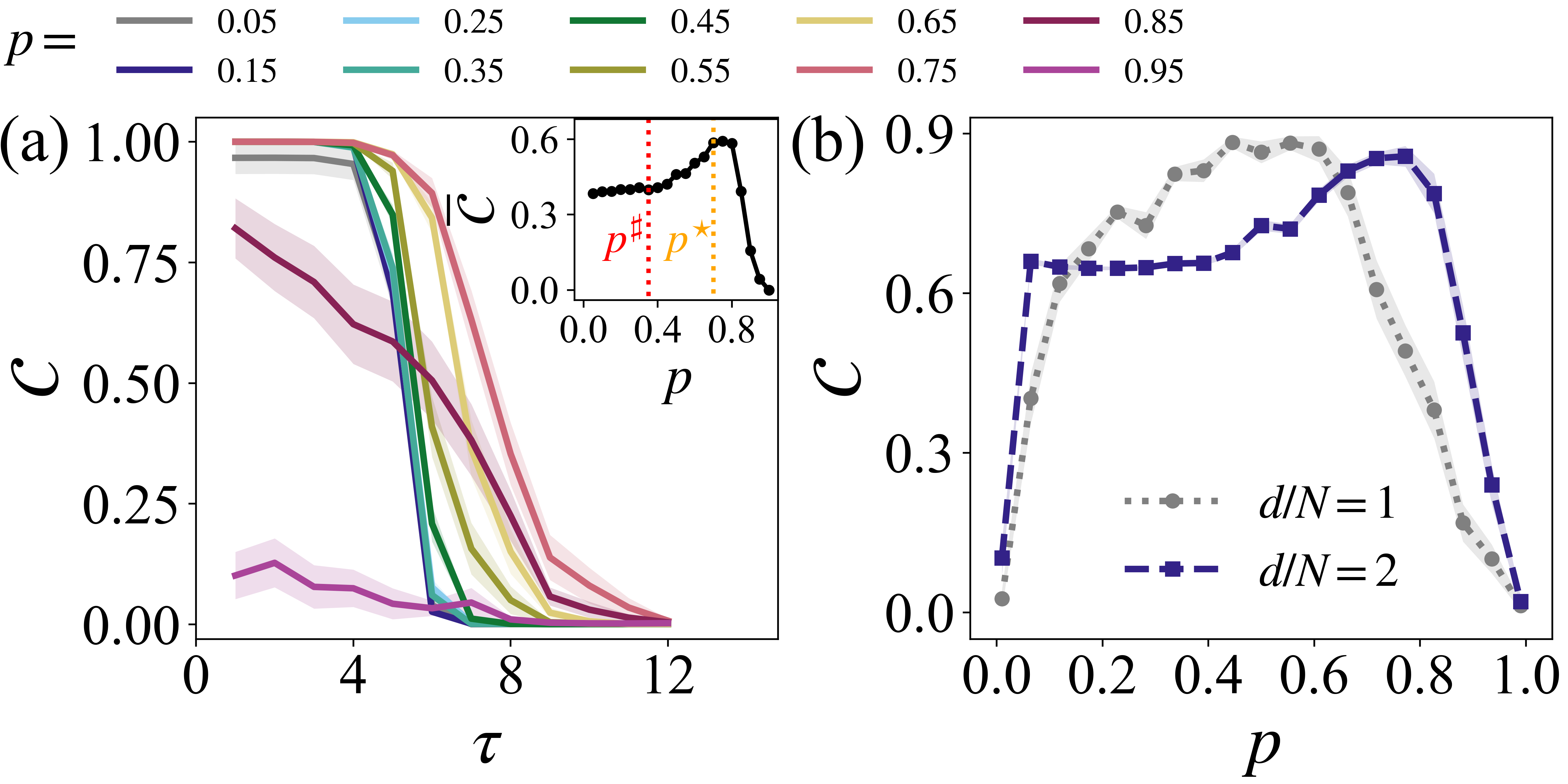}
    \caption{\textbf{Linear memory and nonlinear learnability.} \textbf{(a)} Linear memory as function of delay $\tau$ for various $p$ with $(N,d/N)=(10,2)$. Inset shows the mean memory $\overline{\mathcal{C}}$ over the interval $1\leq\tau\leq12$. \textbf{(b)} Nonlinear learnability as function of $p$ for $d/N=1,2$, with $(\tau,N)=(10,10)$. Each input sequence has $\approx2000$ steps; we discard the first $\approx 500$ (washout) and use the rest for training and testing via supervised linear regression on standardized features (reservoir observables). Each data point is obtained by averaging the performance metrics over $50-100$ random configurations and input sequences. }
    \label{fig4:mem}
\end{figure}

Short-term linear memory is assessed by reconstructing delayed targets $y_n=\theta_{n-\tau}$. Memory at delay $\tau$ is quantified by $\mathcal{C}_{\tau}=\texttt{cov}(\mathbf{y},\bar {\mathbf{y}})^2/\texttt{var}(\mathbf{y}) {\texttt{var}}(\bar {\mathbf{y}})$, where $\bar {\mathbf{y}}$ denotes the array of predictions, and $\texttt{cov}$ and $\texttt{var}$, denote covariance and variance, respectively. Inputs are uniformly distributed in the interval $\theta\in[0,1]$ and rescaled by $10^{-3}$. Results presented in Fig.~\ref{fig4:mem}(a) show that the circuit exhibits excellent memory for intermediate values of $p^{\sharp}\lessapprox p\lessapprox p^{\star}$. Notably, for those parameters neither entanglement nor MM are maximal. Inline with the recent studies on quantum Hamiltonians~\cite{kobayashi2025edge,PhysRevLett.127.100502,gq9r-d5q8}, the reservoir’s overall learnability peaks near the boundary of chaotic/ergodic and integrable/localized regimes. This strongly suggests that the behavior is universal, in the sense that it is insensitive to the details of the underlying dynamical model and the exact mechanism driving the crossover/transition. Thus the proper operating regime is unambiguously diagnosed via the spectral properties of entanglement and long-range magic. While the precise optimum can be task-specific, as in classical reservoir computing~\cite{yildiz2012re, carroll2020reservoir}, peak utility in an intermediate order–chaos regime mirrors behavior seen across complex information-processing and biological systems~\cite{RevModPhys.90.031001, mora2011biological, langton1990computation}.

To expand the analysis of performance, we also consider a widely-studied benchmark, which is the emulation of a nonlinear auto-regressive moving average dynamics. The $\tau$-th order equation is expressed as $y(n+1) = \alpha\,y_n + \beta\,y_n\!\sum_{j=0}^{\tau-1} y_{n-j}+ \gamma \theta_{n-\tau+1}\,\theta_n + \delta,$ where $y_n$ is the $n$-th target and $(\alpha, \beta, \gamma, \delta)\!=\!(0.3, 0.05, 1.5, 0.1)$. The input signal $\theta_n$ is given by the series $\theta_n= 0.1\left(1 + \prod_{x}\sin(\omega x\,n)\right)$, where ${x \in \{\bar{\alpha},\,\bar{\beta},\,\bar{\gamma}\}}$, $\omega=2\pi/T$, and constants are chosen as \((\bar{\alpha}, \bar{\beta}, \bar{\gamma}, T) = (2.11,\, 3.73,\, 4.11,\, 100)\)~\cite{PhysRevApplied.8.024030, suzuki2022natural, atiya2000new}. As shown in Fig.~\ref{fig4:mem}(b), the nonlinear learnability exhibits a finite jump for any $p>0$ and then grows to a maximum near the spectral crossover. Consistent with earlier results for the entanglement-spectrum, the peak performance occurs at different values of $p$ for different depths.

\textbf{\emph{Scalability and anti-flatness.}}---We now discuss the feasibility of accurate information processing in the limit $N\to\infty$. In fast-scrambling circuits~\cite{PRXQuantum.2.030316, ippoliti2022solvable}, expectation values of local observables converge toward ensemble-typical values, and their long-time fluctuations may shrink exponentially with system size~\cite{srimahajariyapong2025connecting, mele2024introduction,cerezo2022challenges}. As a result, in practical situations, the number of measurements required to resolve expectation values and distinguish different input states may also grow exponentially~\cite{thanasilp2024exponential,sannia2025exponential,xiong2025role,sarkar2025concentration,mccaul2025free,xiong2025role}. We address this with via the scaling of anti-flatness of a state~\cite{PhysRevA.109.L040401}, defined as
\begin{align}
    \mathcal{F}(\hat\rho):={\mathrm{Tr}}\hat\rho_R^3-({\mathrm{Tr}}\hat\rho_R^2)^2 ,
\end{align}
which is the variance of a reduced pure state with respect to itself; $\mathcal{F}(\hat\rho)=\mathrm{var}_{\hat\rho_R}(\hat\rho_R)$ where $\hat\rho_{R}=\mathrm{Tr}_{M}\hat\rho$. Similar to nonlocal magic, one has $\mathcal F=0$ for both stabilizer states (entangled or not) and random product-states (which only have local magic)~\cite{cao2024gravitational}. For an equal bipartition and in the absence of symmetries, the ensemble-averaged anti-flatness of Haar-random states satisfies $\mathcal{F}_{\mathrm H}=\mathbb {E}_{\mathrm H}[\mathcal{F}(\hat\rho_{\mathrm H})]\!\sim\!2^{-N}$~\cite{y9r6-dx7p}, the typical behavior of a sufficiently random (nonintegrable) quantum dynamics. 

\begin{figure}[t]
    \centering
    \includegraphics[width=\linewidth]{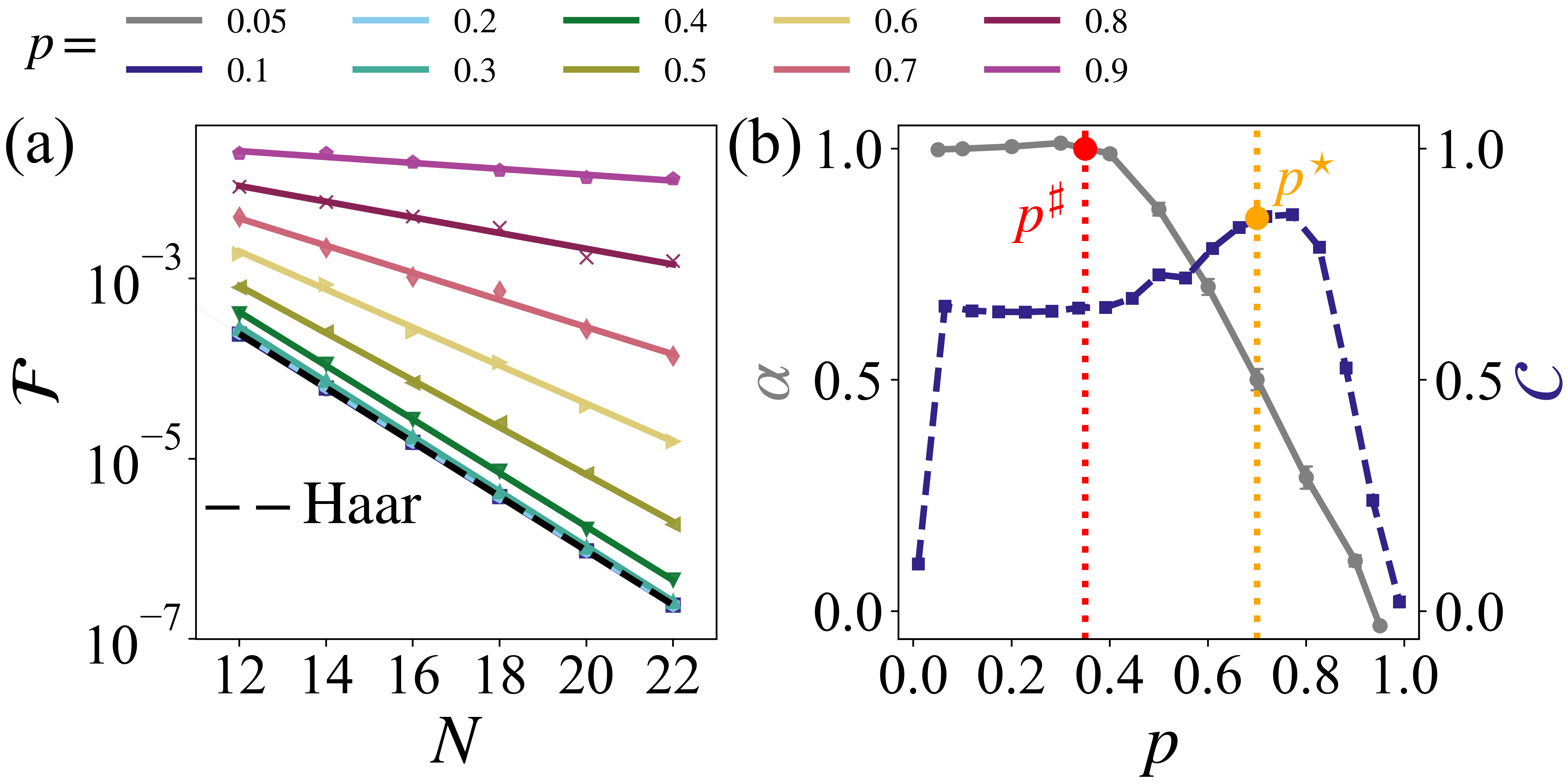}
    \caption{\textbf{Scaling of anti-flatness.} \textbf{(a)} Size-scaling of anti-flatness and linear fits, plotted in logarithmic scale for various $p$. \textbf{(b)} The decay slope $\alpha(p)$, extracted from a linear fit to $\log_2(\mathcal{F})\!\sim -\alpha N$. The results are averaged over $200$ realizations. On the right axis we have replotted the nonlinear memory for the same depth $d=2N$ to match the performance to the obtained points $p^{\sharp}$ and $p^{\star}$.}
    \label{fig5:flatness}
\end{figure}

Demonstrated in Fig.~\ref{fig5:flatness}, one can continuously control the behavior of flatness by tuning $p$ and infer the concentration of measures in the reservoir. Writing $\log_2(\mathcal{F})=-\alpha N+c$, we may interpret $ \alpha\equiv\alpha(d,p)$ as \textit{concentration rate}, where for Haar-random states $\alpha_{\mathrm{H}}=1$. In the limit $N\to\infty$, such states admit reduced subsystems that approach maximally-mixed state $\sim\mathbb{I}/2^N$ and are locally flat ({\it i.e.}, information is ``hidden" from local observers)~\cite{PhysRevLett.109.040502,PhysRevLett.125.030505}. This implies that the expected optimal regime in terms of both performance and scalability may only be achieved when $\mathcal F/\mathcal {F}_{\mathrm{H}}\gg1$ and $0\ll\alpha<\alpha_{\mathrm H}$ (see Fig.~\ref{fig5:flatness}(b)). A weak decay $\alpha\ll\alpha_{\mathrm H}$ (strictly when $\mathcal F\neq0$) signals a weakly entangled (localized) nonstabilizer state. Taken together, the relative value of the anti-flatness and its scaling discriminate physical phases, and are a proxy for both the computational performance and practical readout cost in the large-$N$ limit.

\textbf{\emph{Conclusions and outlook.}}---In this work we introduce a PQR computer with a single control parameter for robust temporal quantum learning. By quantifying \rev{ensemble-typical intrinsic} learnability via the circuits' distance to a quantum chaotic spectral-law, we have shown that \rev{the averaged performance} peaks \rev{\textit{in proximity to a quantum universal entanglement spectrum}}, where nonlocal qunatum resources, namely entanglement and mutual nonstabilizerness, are extensive yet submaximal. Our findings suggest that for QRs targeting nontrivial temporal learning tasks requiring both long intrinsic memory and rich nonlinear processing, only this intermediate regime may provide the required balance between quantum expressivity and operational accessibility. Same principles carry over to other tasks such as the prediction of a chaotic input series~\cite{xia2022reservoir}. \rev{Given} the ``second law of quantum complexity"~\cite{PhysRevD.97.086015}, the gap to maximal complexity (``uncomplexity") \rev{is} a consumable resource for directed computations. Accordingly, dynamics that drive states rapidly to near-Haar typicality (maximally-scrambling) may in general be suboptimal for learnability.

The family of brickwork circuits studied here provide tight control over dynamics, and enable QML architectures grounded in simple physical principles. Our findings however are largely independent of the specific quantum dynamics considered here. The robustness and practicality of our approach can be readily evaluated in both noisy and symmetric settings~\cite{ivaki2025noise, domingo2023taking, trigueros2025nonstabilizerness}. The findings are also relevant to other learning frameworks, particularly quantum kernel methods and extreme learning~\cite{thanasilp2024exponential, xiong2025fundamental}, where the same expressivity-induced concentration is observed. As established, a near flat spectrum for the reduced density matrix points to insensitivity to input data series. This motivates a resource-theoretic formulation of quantum learning based on deviations from thermodynamic flatness~\cite{RevModPhys.91.025001}, which naturally connects to classical simulability at the level of both dynamics and observables. \rev{In particular, assuming noiseless dynamics, it is an open question to what extent some observables around the optimal point may remain classically simulable in an average-case sense~\cite{lh6x-7rc3, rudolph2025pauli}.}

For experimental realizations, one can imagine a conceptually identical setup with random single-qubit Clifford and  parameterized conditional gates (tailored to a device's native gate set), using the well-known Cartan decomposition~\cite{varikuti2025impact}. This allows one to formalize a predictive link between learnability and the joint profile of entangling and nonstabilizing powers of a circuit.

\textbf{\emph{Acknowledgments.}}---This work was supported by the European Union and the European Innovation Council through the Horizon Europe project QRC-4-ESP (Grant Agreement No. 101129663), and EU Horizon Europe Quest project
(No. 10116088), and the Academy of Finland through its QTF Center of Excellence program (Project No. 312298).

\textbf{\emph{Data and code availability.}}---The codes supporting the findings of this work are openly available~\cite{codes}. All data underlying the figures can be reproduced from the released codes.

\section{\bf{Appendix}}

\emph{\textbf{Details on the construction of the reservoir.}}---In reservoir computing, temporal learning is implemented by driving a fixed dynamical system, ``the reservoir", with a sequence of inputs, and training only a final linear readout. The key idea is that the driven quantum dynamics implements a high–dimensional feature map of the input history, so that nonlinear and time-dependent input–output relations can be approximated by simple linear regressions on a set of suitably chosen observables of the system.

In the present work, the reservoir is realized by an $N$–qubit brickwork circuit. At each discrete time step $n$ a scalar input $\theta_n$ is encoded via local rotations on the memory qubits, followed by layers of the probabilistic circuit with a fixed random template. Denoting the reservoir state after step $n$ by $\hat\rho_n$, the driven dynamics can be written as \(\hat\rho_{n+1} = \Phi_{\theta_n}[\hat\rho_n]\), where $\Phi_{\theta_n}$ is a completely positive trace-preserving map. The qubits are partitioned into memory and readout subsets; after each step we measure a set of observables $\{\hat O_\alpha\}_{\alpha=1}^{N_{\mathrm{f}}}$ supported on the readout, obtaining classical features \(x_{n,\alpha} = \mathrm{Tr}(\hat O_\alpha \hat\rho_n)\), $\mathbf{x}_n = (x_{n,1},\dots,x_{n,N_{\mathrm{f}}})^{\mathsf T}$. Since the evolution is contractive on the readout (measure–and–reset), $\mathbf{x}_n$ depends only on a finite window of past inputs $(\theta_{n},\theta_{n-1},\dots)$ set by the intrinsic memory of the circuit. In this work we collect the features only at the end of each encode-and-evolve cycle for a chosen depth $d/N$ (no time-multiplexing).

As mentioned in the text, a suitable reservoir has to satisfy a convergence property, and some initial number of steps, determined by the reservoir dynamics, must be excluded from the classical training. This can be captured, for instance, by monitoring the evolution of an averaged measure of distance between two random mixed states $\hat{\rho},\hat{\rho}'$ under the reservoir dynamics, \textit{e.g.}, $\mathcal{D}\propto \mathrm{Tr}\,|\hat{\rho}-\hat{\rho}'|$, where $|\rho|=\sqrt{\rho^{\dagger}\rho}$. The behavior of $\mathcal{D}$ as a function of the number of iterations $n$ reveals the degree of contractivity of the quantum map and is closely related to the intrinsic \textit{computational} memory of the system. For a Markovian channel, such as the one considered here, $\mathcal{D}$ typically decays exponentially as $\mathcal{D}(n,\eta)=\mathcal{D}_0\,e^{-n\eta}$, where $\eta\equiv\eta (p,d,\mathcal{N})$ is the decay rate, and $\mathcal{N}=N_R/N_M$ indicates the relative spatial size of memory (hidden) and readout (accessible) subsets. When $\mathcal{N}\!\gg\!1$, the reservoir hosts little hidden capacity and memory decays quickly. Conversely, for $\mathcal{N}\!\ll\!1$, the decay is slow, but the accessible readout captures little of the stored information, greatly limiting the usable feature space. We fix $\mathcal{N}=1$ and mainly focus on the role of $p$.

For each task we specify a target time series $\{y_n\}$, which is some functional of the input history, \textit{e.g.}, $y_n = \theta_{n-\tau}$ for linear memory at delay $\tau$, or a nonlinear combination of past inputs for the benchmark task considered in the main text. After discarding the first $T_{\mathrm{w}}$ “washout’’ steps to remove dependence on the initial state, we collect a training set of length $T_{\mathrm{tr}}$ and stack the features into a design matrix \(X \in \mathbb{R}^{T_{\mathrm{tr}}\times N_{\mathrm{f}}}\), $X_{n\alpha} = \tilde x_{n,\alpha}$, where $\tilde x_{n,\alpha}$ denotes standardized features (zero mean and unit variance over the training set for each $\alpha$) to improve numerical stability. The corresponding target vector is $\mathbf{y} \in \mathbb{R}^{T_{\mathrm{tr}}}$ with entries $y_n$. The readout is a linear map $\bar y_n = \mathbf{w}^{\mathsf T}\mathbf{x}_n$ with trainable weights $\mathbf{w}\in\mathbb{R}^{N_{\mathrm{f}}}$. We determine $\mathbf{w}$ via ridge (Tikhonov) regression, by minimizing the regularized quadratic loss \(\mathcal{L}(\mathbf{w}) = \frac{1}{T_{\mathrm{tr}}}\,\|X\mathbf{w}-\mathbf{y}\|_2^2 + \lambda \|\mathbf{w}\|_2^2,\) where $\lambda>0$ is a small regularization parameter that suppresses overfitting. The minimizer has the closed form $\mathbf{w}_* = (X^{\mathsf T}X + \lambda I)^{-1}X^{\mathsf T}\mathbf{y}$ and in our calculations we set $\lambda=10^{-3}-10^{-4}$.

After training, performance is evaluated on independent test sequences with the same fixed reservoir realization and readout weights. The averaged memory capacity and nonlinear learnability metrics reported in the main text are then computed from the Pearson correlation between $\{y_n\}$ and $\{\bar y_n\}$ for independent input sequences and random reservoir realizations. In practice, no direct knowledge of the full wavefunction is required at any stage, and the probabilistic quantum system is never optimized internally during training as all trainable parameters reside exclusively in the classical readout layer.

\emph{\textbf{On the relation between $p^{\star}$ and $p^{\sharp}$.}}---Here we establish a connection between $p^{\star}$, which is associated with the behavior of the entanglement spectrum, and $p^{\sharp}$, which characterizes the dynamical growth of entanglement entropy. The latter, in a generic random circuit, is also closely related to the behavior of mutual magic and the scaling of anti-flatness. To this end, we consider a diluted Haar model. In this effective picture, states are locally transformed as
\[\hat{\rho}\mapsto \begin{cases}\hat{\rho}, & \text{w.p. }p; \\ \hat{U}_{\mathrm{H}}\,\hat{\rho}\,\hat{U}_{\mathrm{H}}^\dagger, & \text{w.p. } 1-p, \end{cases}\]
where the initial state $\hat{\rho}$ is assumed to be a random product state, and $\hat{U}_{\mathrm{H}}$ is a two-qubit unitary sampled uniformly at random from the Haar measure. Although the correspondence between this model and the one in the main text is not exact, for sufficiently large system size and depth both exhibit a crossover between quantum chaotic and integrable dynamics around the same values of $p$.

There are two \textit{speed limits} that directly control the appearance of crossover points, namely, entanglement velocity $v_{\mathrm{E}}$, and butterfly velocity $v_{\mathrm{B}}$. The former controls the growth of state entanglement entropy, while the latter controls operator spreading and the dynamics of out-of-time-order correlators~\cite{PhysRevB.100.125115, PhysRevLett.123.190602, PhysRevX.8.021014, roberts2015localized}. Under dilution, the operator front, associated with the spread of total weights on Pauli strings, performs a "lazy" biased random walk, so in a coarse-grained picture we expect $v_{\mathrm{B}}(p)\approx (1-p) v_{\mathrm{B}}(0)$~\cite{PhysRevX.8.021013}, with $v_{\mathrm{B}}(0)=3/5$ the velocity in a circuit consisting of local Haar-random unitaries acting on qubits. For generic chaotic systems there are substantial evidences that show $v_{\mathrm{B}} \geq v_{\mathrm{E}}$~\cite{PhysRevX.8.021013, jonay2018coarse}. This implies that a system may exhibit a universal entanglement spectrum, as predicted by random matrix theory, even while quantities such as the entanglement entropy have not yet reached their long-time thermal values~\cite{PhysRevLett.123.190602, PhysRevB.100.125115}. This is reasonable, since the spectral statistics as characterized by gap ratio $r$ primarily takes into account the local spectral structure and level repulsion between nearby eigenvalues. Operator hydrodynamics further yields a relation between the \emph{$S_2$} entanglement velocity and the butterfly velocity~\cite{PhysRevX.8.021013}
\[
v_{\mathrm{E}}=\frac{\log(1-v_{\mathrm{B}}^{2})}{\log[({1-v_{\mathrm{B}})}/({1+v_{\mathrm{B}}})]}.
\]
On the interval $v_{\mathrm{B}}\in[0,3/5]$ this relation is close to linear (to within a few-percent accuracy), giving $v_{\mathrm{E}}(0)\approx 0.53\,v_{\mathrm{B}}(0)$. In numerics (for the von Neumann entropy) one finds that $v_{\mathrm{E}}(p)$ is well approximated by a linear decrease with $p$ and we may write $v_{\mathrm{E}}(p)\approx (1-p) v_{\mathrm{E}}(0)$.

Given these, for diluted circuits, we may estimate $p^{\sharp} \approx 1 - {N}/[{2 v_{\mathrm{E}}(0) d}]$, which follows from requiring that $(1 - p^{\sharp}) v_{\mathrm{E}}(0) d \simeq N/2$. In the limit $d/N \ll 1$ (\textit{e.g.}, $d \sim \mathcal{O}(\log N)$), one has $p^{\sharp} < 0$, which in practice means the system cannot reach the entanglement-saturated ("fully chaotic") regime within that depth. In the opposite limit $d/N \gg 1$, one instead obtains $p^{\sharp} = 1$, meaning the crossover to the integrable regime is pushed to the trivial endpoint, as expected. Similarly, one finds $p^{\star} \approx 1 - {N}/[{2 v_{\mathrm{B}}(0) d}]$, and since $v_{\mathrm{B}}(0)>v_{\mathrm{E}}(0)$, it follows that $p^{\sharp}(d) \leq p^{\star}(d)$, a signature of multi-step thermalization. It also follows that optimal learning around $p^{\star}$ signifies the point where the operator front has just touched the boundary of the circuit. Generally for $0<(p^{\sharp},p^{\star})<1$ one finds
\[
\frac{1-p^{\sharp}}{1-p^{\star}}\approx\frac{v_{\mathrm{B}}(0)}{v_{\mathrm{E}}(0)},
\]
which agrees well with the numerical calculations. The crossover points can be evaluated as the smallest value of $p$ at which a measure $f(p)$ departs appreciably from its plateau value.
Specifically, one can estimate the plateau mean $\mu$ in a range where the curve is approximately flat (excluding the initial transient), and identify the first $p$ such that \({|f(p)-\mu|}/{|\mu|}>\varepsilon\) within a fixed tolerance $\varepsilon$ (\textit{e.g.},\ $\varepsilon=0.01-0.02$).
This relative criterion allows consistent comparison across measures with different overall scales. The resulting estimates are additionally limited by the discrete resolution of the $p$ grid.

\bibliography{ref.bib}

\end{document}